\documentclass[a4,journal,comsoc]{IEEEtran}

\def\ba{ \begin{eqnarray} }
\def\ea{ \end{eqnarray} }
\def\be{\begin{equation} }
\def\ee{\end{equation} }
\def\bef{ \begin{figure} }
\def\eef{ \end{figure} }
\def\bc{ \begin{center} }
\def\ec{ \end{center} }

\begin{document}
\input psfig
\date{\today}


\title{Modify-and-Forward for Securing Cooperative Relay Communications}
\author{Sang Wu Kim\\
Department of Electrical and Computer Engineering\\
Iowa State University\\
Ames, IA 50011\\
E-mail: swkim@iastate.edu}

\maketitle

{\it \bf Abstract} - We proposed a new physical layer technique that can enhance the security of cooperative relay communications. The proposed approach modifies the decoded message at the relay according to the unique channel state between the relay and the destination such that the destination can utilize the modified message to its advantage while the eavesdropper cannot.  We present a practical method for securely sharing the modification rule between the legitimate partners and present the secrecy outage probability in a quasi-static fading channel.  It is demonstrated that the proposed scheme can provide a significant improvement over other schemes when the relay can successfully decode the source message.

\section{Introduction}
In recent years, there have been considerable efforts devoted to using the channel to provide security in wireless communications.  It is shown in \cite{s2} that fading alone guarantees that information-theoretic security is achievable, even when the eavesdropper has a better average SNR than the legitimate receiver. 
A traditional approach to enhancing the secrecy rate is to introduce interference (jamming) into the channel so as to harm the eavesdropper's ability to eavesdrop while strengthening the ability for legitimate entities to communicate.  This idea has appeared in the literature under the name of artificial noise \cite{a3}, cooperative jamming (CJ) \cite{ja1,ja2,ja5,a33,gabry}, or noise forwarding (NF) \cite{ja6,ulu1}.

In this paper we propose a new physical layer technique that  can enhance the security of cooperative relay communications.  Unlike traditional approaches in which no context (message) is sent by the relay, in the proposed scheme the relay decodes the source message $X$ and forwards a {\it modified} message $X'$ to the destination such that the intended destination can utilize $X'$ to its advantage while the eavesdropper cannot.   The basic idea is to exploit the unique physical channel state between the relay and the destination as the inherent shared secret in sharing $X'-X$ without exchanging any information about $X'-X$.   Once the difference $X'-X$ is known at the destination, it can be canceled from the modified message $X'$ to get the original message $X$, while the eavesdropper without knowing the difference\footnote{The eavesdropper cannot determine the physical channel state between the legitimate nodes  as long as the former is more than half of the wavelength away from the latter.} cannot extract $X$ from $X'$.  The additional information about $X$ provided by the relay can improve the rate towards the intended destination without improving the rate towards the eavesdropper.  Hereafter, the proposed scheme will be referred to as {\it modify-and-forward} (MF).

We present a practical method for securely sharing the difference $X'-X$ (or modification rule in general) by exploiting the unique physical channel state between the legitimate partners.  We characterize the security level in a quasi-static fading environment by computing the secrecy outage probability that provides the fraction of fading realizations for which the wireless channel cannot support a target secure rate.   We compare the secrecy outage probability of the proposed scheme with that of direct transmission (DT), decode-and-forward (DF), and CJ under different system setups.

\section{System Model}
\begin{figure}[h!]
\centerline{\psfig{figure=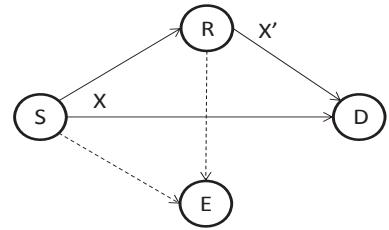,width=5cm,height=3cm}}
\caption{Cooperative relay communication model for modify-and-forward relaying.}
\label{fig0}
\end{figure}

We consider the cooperative relay communication system shown in Fig. \ref{fig0} in which a source (S) communicates with a destination (D) with the help of a relay (R) in the presence of a eavesdropper (E).    We assume that each node carries a single omnidirectional antenna.  Channels between all pairs of nodes are modeled as independent quasi-static Rayleigh fading channels: fading coefficients remain constant during the transmission of an entire codeword but they change from one codeword to another according to a complex Gaussian distribution.

In the first phase, S broadcasts the message $X$ to D and E.  In the second phase, the relay decodes the message transmitted by S, modifies the decoder output to $X'$ and broadcasts $X'$ to D and E.  We require the relay to fully decode the source message $X$ and the source to remain silent during the second phase.   We assume that $X$ and $X'$ are of length $n$ and are independently chosen from a Gaussian random codebook of $M$ codewords.  We also assume that each codeword is chosen with equal probability and that $E[X]=E[X']=0$ and $E[||X||^2]/n = E[||X'||^2]/n =P$.  Thus the total transmission power is $2P$.

The received signals at the destination that are originated from the source and relay are, respectively, given by
    \ba
    Y_{sd} & = & h_{sd} X+N_{sd} \label{ysd}\\
    Y_{rd} & = & h_{rd} X'+N_{rd} \label{yrd}
    \ea
where  $h_{ij}$ is the channel gain between the node $i$ and node $j$, and $N_{ij}$ is white Gaussian noise with mean zero and variance $\sigma_n^2$.
Once $X'-X$ is known at the destination, it can be removed from $Y_{rd}$ to get
    \ba
    Y_{rd}' & = & Y_{rd}-h_{rd} (X'-X)\\
    & = & h_{rd} X+N_{rd}
    \ea
and $X$ can be decoded based on $Y_{sd}$ and $Y_{rd}'$.

We assume that the eavesdropper knows that the message is modified by the relay.  However, without knowing the difference $X'-X$, it has to discard the signal received from the relay $Y_{re}=h_{re} X' +N_{re}$ and decode $X$ based on the signal received from the source only:
    \be
    Y_{se}=h_{se} X+N_{se} \label{ye}
    \ee
where $h_{se}$ is the channel gain between the source and the eavesdropper and $N_{se}$ is the noise.   This is because $Y_{re}$ does not provide any information about $X$ unless $X'-X$ is known.

The question is how to achieve the agreements on message modification secretly between the relay and the destination. Only when two nodes share the same modification rule they can achieve high secrecy rate.  Our approach is based on the uniqueness and reciprocity of wireless fading channel.  The
reciprocity theory demonstrates that bidirectional wireless channel states should be identical between two transceivers during the channel's coherence time \cite{b2}. We use this unique channel state as the inherent shared secret between the relay and the destination for message modification and restoration.  As long as the eavesdropper is more than half of the wavelength away from legitimate communicators, the channel states he observed should be independent to the channel state between the legitimate ones \cite{b3}. This means the eavesdropper can never eavesdrop the secret $X'-X$ shared between legitimate communicators. Since the legitimate communicators do not exchange any information about $X'-X$, our approach provides a strong security.  The uniqueness of the wireless channel between two locations has also been utilized in authenticating legitimate users \cite{green}.

\section{Secrecy Outage Probability}
In this section we derive the secrecy outage probability which provides the fraction of fading realizations for which the wireless channel cannot support a target secrecy rate of $R$.  It provides a security metric for the situation where
the source and destination have no channel state information about the eavesdropper.

\subsection{Modify-and-Forward}
The maximum rate at which the relay and the destination can reliably decode the message $X$ is given by \cite{laneman}
    \ba
    C_d & = & \min \left\{ \frac{1}{2} \log_2 \left( 1+ |h_{sr}|^2  P/\sigma_n^2 \right), \right.\nonumber \\
     & & \left. \frac{1}{2} \log_2 \left( 1+ (|h_{sd}|^2  +  |h_{rd}|^2 ) P/\sigma_n^2 \right) \right\} \label{cd1}
    \ea
where the factor $1/2$  accounts for the two-phase transmission.  Similarly, the maximum rate at which E can reliably decode the message $X$ is
    \be
    C_e = \frac{1}{2}\log_2 \left( 1+ |h_{se}|^2   P/\sigma_n^2 \right) \label{ce}
    \ee
because the eavesdropper cannot utilize the modified message which is sent by the relay.  Then, the instantaneous secrecy capacity between S and D is \cite{s1}
    \be
    C_s=\max( C_d-C_e,0) \label{cs}
    \ee

Communication is secure if the instantaneous secrecy capacity  $C_s$ is higher than the target secrecy rate $R$ (b/s/Hz).  If $C_s<R$, then security is compromised and secrecy outage occurs.  The secrecy outage probability for the proposed scheme can be shown to be
    \ba
    P_o (R) & = & P(C_s < R) \\
    & = & 1-\frac{1}{\gamma_{rd}-\gamma_{sd}} \left(1+\frac{\gamma_{rd}}{\gamma_{sr}} \right) e^{-(2^{2R}-1)\left(\frac{1}{\gamma_{sr}}+\frac{1}{\gamma_{rd}} \right)} \nonumber\\
    & & \times \left[ \frac{1}{\frac{1}{\gamma_{sr}}+\frac{1}{\gamma_{rd}}} - \frac{1}{\frac{2^{-2R}}{\gamma_{se}}+\frac{1}{\gamma_{sr}}+\frac{1}{\gamma_{rd}}} \right] \nonumber\\
    & & + \frac{1}{\gamma_{rd}-\gamma_{sd}} \left(1+\frac{\gamma_{sd}}{\gamma_{sr}} \right) e^{-(2^{2R}-1)\left(\frac{1}{\gamma_{sr}}+\frac{1}{\gamma_{sd}} \right)} \nonumber\\
    & & \times \left[ \frac{1}{\frac{1}{\gamma_{sr}}+\frac{1}{\gamma_{sd}}}  -\frac{1}{\frac{2^{-2R}}{\gamma_{se}}+\frac{1}{\gamma_{sr}}+\frac{1}{\gamma_{sd}}}\right] \label{mf}
    \ea
where  $\gamma_{sd}  =  E[|h_{sd}|^2] P/\sigma_n^2$,
$\gamma_{rd}  =  E[|h_{rd}|^2] P/\sigma_n^2$,
$\gamma_{se}  =  E[|h_{se}|^2] P/\sigma_n^2$, and
$\gamma_{re}  =  E[|h_{re}|^2] P/\sigma_n^2$.  Proof of (\ref{mf}) is provided in Appendix A.

\subsection{Direct Transmission}
For the direct transmission (DT), within a transmission slot, the source transmits its $n$ encoded symbols directly to the destination using  the available transmit power of $2P$.   The secrecy outage probability with the DT is given by \cite{s2}
    \ba
    P_{o}(R) & = & 1-\frac{\gamma_{sd}}{\gamma_{sd} + 2^{R}\gamma_{se}} \exp \left( -\frac{2^{R}-1}{2\gamma_{sd}}\right) \label{dt}
    \ea
where the factor $2$ in front of $\gamma_{sd}$  accounts for the total transmit power of $2P$.

\subsection{Decode-and-Forward}
Like MF, decode-and-forward (DF) is also a two-phase scheme.  The first phase is the same as in the MF scheme.  In the second phase, the relay decodes the information transmitted by the source and re-encodes it using the same codeword as the source to transmit the information to D.  Thus the total transmission power is $2P$.
The secrecy outage probability with the DF is given by \cite{gabry}
    \ba
    P_{o}(R) & = & \frac{a(\gamma_{re})-a(\gamma_{se})}{\gamma_{re}-\gamma_{se}}\nonumber\\
    & & \hspace{-5mm} +\frac{\gamma_{sr}2^{-2R} a(\gamma_{se}) (h(\gamma_{se},\gamma_{sd})-h(\gamma_{se},\gamma_{rd}))}{(\gamma_{re}-\gamma_{se})(\gamma_{rd}-\gamma_{sd})}\nonumber \\
    & & \hspace{-5mm} -\frac{\gamma_{sr}2^{-2R} a(\gamma_{re}) (h(\gamma_{re},\gamma_{sd})-h(\gamma_{re},\gamma_{rd}))}{(\gamma_{re}-\gamma_{se})(\gamma_{rd}-\gamma_{sd})}
    \ea
where
    \ba
    h(x,y) & = & \frac{\gamma_{sr}}{x(1+\gamma_{sr}/y)+\gamma_{sr}2^{-2R}}\\
    a(x) & = & \frac{x^2}{\gamma_{sr}2^{-2R}+x} \exp\left(-\frac{2^{-2R}-1}{x} \right)
    \ea

\subsection{Cooperative Jamming}
Various cooperative jamming (CJ) schemes that involve the transmission of jamming signals from different nodes have been proposed \cite{ja1,ja2,a33}.  In this paper we consider the cooperative jamming scheme where, while S transmits, the relay transmits a jamming signal that is independent of the source message with the purpose of confounding E. The jamming signal, white Gaussian noise, causes interference at both D and E. The total transmission power\footnote{The total transmission power of CJ schemes in \cite{ja2,a33} is $3P$ because each of three nodes (source, relay, and destination) transmits with power $P$.} is $2P$ as the source and relay transmits with power $P$.
The secrecy outage probability for the CJ is given by \cite{ja1}
    \ba
    P_{o}(R) & = & 1-\frac{2^{-\kappa}}{\gamma_{rd}\gamma_{re}}\frac{\gamma_{re}}{\left(\kappa+\frac{1}{\gamma_{rd}}-\frac{\beta}{\gamma_{re}}\right)}\nonumber\\
    & & + \frac{2^{-\kappa}}{\gamma_{rd}\gamma_{re}} \left( \kappa+\frac{1}{\gamma_{rd}}-\frac{\beta}{\gamma_{re}} \right)^{-2}\nonumber\\
    & & \times \left[ \beta \left( \kappa+\frac{1}{\gamma_{rd}}-\frac{\beta}{\gamma_{re}} +1 \right) \Omega\left(\frac{1+\beta}{\gamma_{re}}\right) \right. \nonumber\\
    & & \left. +\left( \kappa+\frac{1}{\gamma_{rd}}-\frac{\beta}{\gamma_{re}} -\beta \right) \right.\nonumber\\
    & & \times \left. \Omega \left( \frac{1+\beta}{\beta} \left( \kappa+\frac{1}{\gamma_{rd}} \right)\right) \right] \label{cj}
    \ea
where $\kappa=(2^{2R}-1)/\gamma_{sd}$, $\beta=2^{2R}\gamma_{se}/\gamma_{sd}$, and $\Omega(x)=e^x E_1(x)$ where $E_1(x)=\int_x^\infty u^{-1}e^{-u} du$.


\subsection{Numerical Results}
\begin{figure}[h!]
\centerline{\psfig{figure=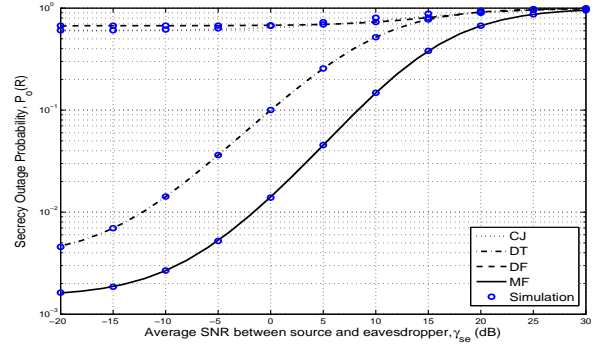,width=9cm,height=5cm}}
\caption{Secrecy outage probability, $P_o(R)$, versus average SNR between source and eavesdropper, $\gamma_{se}$ (dB); $R=0.1$b/s/Hz,  $\gamma_{sd}=10$dB, $\gamma_{sr}=20$dB, $\gamma_{rd}=20$dB, $\gamma_{re}=15$dB.}
\label{fig2}
\end{figure}

Fig. \ref{fig2} shows the secrecy outage probability, $P_o(R)$, versus the average signal-to-noise ratio (SNR) between the source and the eavesdropper, $\gamma_{se}$.  As expected the secrecy outage probability increases with increasing $\gamma_{se}$ because the rate at which the eavesdropper can reliably decode the message increases as the channel condition between the source and itself improves.  It can also be seen that the improvement provided by MF over DF is more significant at lower $\gamma_{se}$.   This is because the eavesdropper relies sorely on the channel between the source and eavesdropper in MF, while in DF the eavesdropper can rely on the channel between the relay and itself when $\gamma_{se}$ is low.  Similarly, in DT the eavesdropper relies sorely on the channel between the source and itself and therefore the secrecy outage probability depends heavily on $\gamma_{se}$.

\begin{figure}[h!]
\centerline{\psfig{figure=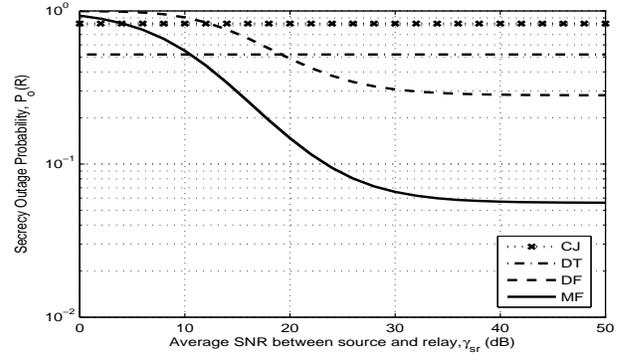,width=9cm,height=5cm}}
\caption{Secrecy outage probability, $P_o(R)$, versus average SNR between source and relay, $\gamma_{sr}$ (dB); $R=0.1$b/s/Hz,  $\gamma_{sd}=10$dB, $\gamma_{se}=10$dB, $\gamma_{rd}=20$dB, $\gamma_{re}=15$dB.}
\label{fig2_1}
\end{figure}

Fig. \ref{fig2_1} shows the secrecy outage probability, $P_o(R)$, versus the average SNR between the source and the relay, $\gamma_{sr}$.  For DF and MF schemes, the relay has to decode the source message in order to provide any additional information to the destination.  Therefore, if $\gamma_{sr}$ is low, the secrecy outage probability for DF and MF is high because the relay cannot decode the source message.  However, if $\gamma_{sr}$ is high enough such that the relay can decode the source message, then it can provide additional information to the destination, which increases the secrecy capacity.     At sufficiently high $\gamma_{sr}$, the secrecy outage probability for DF and MF remains constant because all other channel gains are assumed to be constant.

\begin{figure}[h!]
\centerline{\psfig{figure=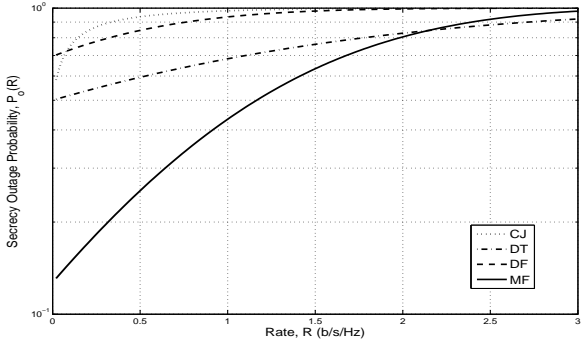,width=9cm,height=5cm}}
\caption{Secrecy outage probability $P_o(R)$ versus rate $R$ (b/s/Hz); $\gamma_{sd}=10$dB, $\gamma_{sr}=20$dB, $\gamma_{rd}=20$dB, $\gamma_{se}=10$dB, $\gamma_{re}=15$dB.}
\label{fig3}
\end{figure}

Fig. \ref{fig3} shows the secrecy outage probability, $P_o(R)$, versus the target secrecy rate $R$.   It can be seen that the improvement that MF provides over the traditional approaches is more significant when the target secrecy rate $R$ is smaller.  However, if $R$ is above a threshold, DT provides the smallest secrecy outage probability, although the secrecy outage probability in that rate region is unacceptably high.  It can also be seen from Figs. \ref{fig2}-\ref{fig3} that MF can always provide a lower secrecy outage probability than DF under any channel conditions and rates.

\section{Conclusion}
We proposed a new physical layer technique that can enhance the security of cooperative relay communications. The proposed approach modifies the decoded message at the relay according to the unique channel state between the relay and the destination such that the destination can utilize it to its advantage while the eavesdropper cannot.  We derived the secrecy outage probability in quasi-static fading channel, and compared with direct transmission, decode-and-forward, cooperative jamming under different system setups.  Numerical results reveal that each scheme provides an advantage over the others depending on the channel gains and secrecy rates, although the proposed scheme can always provide a lower secrecy outage probability than decode-and-forward scheme.  The proposed approach can provide a significant improvement over other schemes when the relay can successfully decode the source message.

\section*{Appendix A}
In this appendix we provide a proof of (\ref{mf}).  Let
    \ba
    X & = & (|h_{sd}|^2+|h_{rd}|^2)P/\sigma_n^2 \label{ap1}\\
    Y & = & |h_{se}|^2 P/\sigma_n^2 \label{ap2}\\
    Z & = & |h_{sr}|^2 P/\sigma_n^2 \label{ap3}
    \ea
Since $h_{ij}$'s are complex Gaussian, $i,j \in \{s,r,d\}$, the probability density function of $X$, $Y$, and $Z$ are given by
    \ba
    f_X(x) & = & \frac{\exp(-x/\gamma_{rd})-\exp(-x/\gamma_{sd})}{\gamma_{rd}-\gamma_{sd}} \label{fx}\\
    f_Y(y) & = & \frac{\exp(-y/\gamma_{se})}{\gamma_{se}} \label{fy} \\
    f_Z(z) & = & \frac{\exp(-z/\gamma_{sr})}{\gamma_{sr}} \label{fz}
    \ea
where $\gamma_{ij}=E[|h_{ij}|^2]P/\sigma_n^2$.   Then,
    \ba
    P_o(R) & = & P(\min \{ \log_2 (1+Z),\log_2(1+X) \} \nonumber \\
    & & \hspace{5mm} < \log_2(1+Y)+2R ) \\
    & = & P(\log_2 (1+\min\{ X,Z\}) \nonumber \\
    & & \hspace{5mm} < \log_2 (1+Y) +2R )\\
    & = & P(2^{-2R} (1+\min \{ X,Z\}) -1 < Y )\\
    & = & P(2^{-2R} (1+X) -1 < Y)P(Z>X) \nonumber \\
    & & + P(2^{-2R} (1+Z) -1 < Y)P(Z<X)
    \ea
If $2^{-2R} (1+X) -1 <0$ or $X<2^{2R}-1$, then $P(2^{-2R} (1+X) -1 < Y)=1$ because $Y>0$.  Similarly, if $2^{-2R} (1+Z) -1 <0$ or $Z< 2^{2R}-1$, then $P(2^{-2R} (1+Z) -1 < Y)=1$.  Therefore, we get
    \ba
    P_o(R) & \hspace{-3mm} = & \hspace{-3mm} \int_0^{2^{2R}-1} f_X(x)  \int_x^\infty  f_Z(z) dz dx \nonumber \\
    & + & \hspace{-3mm} \int_{2^{2R}-1}^\infty f_X(x) \int_{2^{-2R}(1+x)-1}^\infty \hspace{-10mm} f_Y(y) dy \int_x^\infty f_Z(z) dz dx\nonumber \\
    & + & \hspace{-3mm} \int_0^{2^{2R}-1} f_Z(z)  \int_z^\infty  f_X(x) dx dz \nonumber \\
    & + & \hspace{-3mm} \int_{2^{2R}-1}^\infty \hspace{-3mm} f_Z(z) \int_{2^{-2R}(1+z)-1}^\infty \hspace{-8mm} f_Y(y) dy \int_z^\infty \hspace{-3mm} f_X(x) dx dz\\
    & = & \frac{\gamma_{sr}(1+\gamma_{sr})}{(\gamma_{sr}+\gamma_{rd})(\gamma_{sr}+\gamma_{sd})} \nonumber \\
    & & -\frac{e^{-[(2^{2R}-1)/\gamma_{sr}]} (1+\gamma_{sr})2^{-2R}}{(\gamma_{rd}-\gamma_{sd})\gamma_{sr}\gamma_{se}}\nonumber\\
    & & \cdot \left[ \frac{e^{-[(2^{2R}-1)/\gamma_{rd}]}}{\left(\frac{2^{-2R}}{\gamma_{se}}+\frac{1}{\gamma_{sr}}+\frac{1}{\gamma_{rd}}\right)\left( \frac{1}{\gamma_{sr}}+\frac{1}{\gamma_{rd}} \right)} \right. \nonumber\\
    & & \left.  -\frac{e^{-[(2^{2R}-1)/\gamma_{sd}]}}{\left(\frac{2^{-2R}}{\gamma_{se}}+\frac{1}{\gamma_{sr}}+\frac{1}{\gamma_{sd}}\right)\left( \frac{1}{\gamma_{sr}}+\frac{1}{\gamma_{sd}} \right)} \right]
    \ea

\section*{Acknowledgement}
This work was supported in part by the U.S. National Science Foundation under grant ECCS 1254086.


\begin{thebibliography}{99}
\bibitem{s2}
J. Barros and M. Rodrigues, ``Secrecy capacity of wireless channels,"
in {\it Proc. IEEE Int. Symp. Information Theory}, pp. 356–-360, Seattle, WA, Jul. 2006.

\bibitem{a3}
S. Goel and R. Negi, ``Guaranteeing secrecy using artificial noise,"
{\it IEEE Trans. Wireless Commun.}, vol. 7, no. 6, pp. 2180–-2189, Jun.
2008.

\bibitem{ja1}
J.P. Vilela, M. Bloch, J. Barros, and S.W. McLaughlin, ``Wireless secrecy regions with friendly jamming," {\it IEEE Tr. Infor. Forensics and Security}, pp.256--266, VOL. 6, No. 2, Jun. 2011.

\bibitem{ja2}
Z. Ding, Member, K.K. Leung, D.L. Goeckel, and D. Towsley, ``Opportunistic relaying for secrecy communications:
Cooperative jamming vs. relay chatting," {\it IEEE Tr. on Wireless Commun.}, pp. 1725--1729, Jun. 2011.

%

\bibitem{ja5}
L. Dong, Z. Han, A. P. Petropulu, and H. V. Poor, ``Improving wireless
physical layer security via cooperating relays," {\it IEEE Trans. Signal
Process.}, vol. 58, no. 3, pp. 1875–-1888, Mar. 2010.

\bibitem{a33}
J. Huang, A. Mukherjee, and A.L. Swindlehurst, ``Outage performance of amplify-and-forward channels with an unautheticated relay," in {\it Proc. of IEEE ICC}, 2012.

\bibitem{gabry}
F. Gabry, R. Thobaben and M. Skoglund, ``Outage performances for amplify-and-forward, decode-and-forward and cooperative jamming strategies for the wiretap channel," in {\it Proc. of IEEE WCNC}, 2011.

\bibitem{ja6}
L. Lai and H.E.Gamal, ``The relay-eavesdropper channel: Cooperation for secrecy," {\it IEEE Tr. on Infor. Th.}, pp.4005--4019, Sep. 2008.

\bibitem{ulu1}
R. Bassily and S. Ulukus, ``Deaf cooperation and relay selection
strategies for secure communication in multiple relay networks," {\it IEEE
Trans. Signal Process.}, vol. 61, no. 6, pp. 1544--1554, Mar. 2013.
%

%
%
%
%


\bibitem{b2}
R. Wilson, D. Tse, and R. A. Scholtz,``Channel identification: Secret
sharing using reciprocity in ultrawideband Channels," {\it IEEE Tr. Information Forensics and Security}, Sep. 2007.

\bibitem{b3}
W.C.Jakes Jr., {\it Microwave Mobile Communications}, Wiley, 1974.

\bibitem{green}
L. Xiao, L. Greenstein, N. Mandayam, and W. Trappe, ``Fingerprints in the Ether: Using the physical layer
for wireless authentication," in {\it Proc. of IEEE ICC}, 2007.



%
%
%

\bibitem{laneman}
J. N. Laneman and G. W. Wornell, ``Distributed space-time coded protocols for exploiting cooperative diversity in wireless networks," {\it IEEE Trans. Inform. Theory}, vol. 49, no. 10, pp. 2415--2425, Oct. 2003.

\bibitem{s1}
S. Leung-Yan-Cheong and M. Hellman, ``The Gaussian wire-tap
channel," {\it IEEE Trans. Inf. Theory}, vol. 24, no. 4, pp. 451–-456, Jul. 1978.

%


\end{thebibliography}
\end{document}